\documentclass[prl,twocolumn]{revtex4}
\begin{document}

\title{
A Bell-Evans-Polanyi principle for molecular dynamics trajectories and its 
implications for global optimization}

\author{Shantanu Roy, Waldemar Hellmann, Stefan Goedecker}
\affiliation{ Institut f\"{u}r Physik, Universit\"{a}t Basel, Klingelbergstr.82, 4056 Basel, Switzerland \\}

\begin{abstract}
The Bell-Evans-Polanyi principle that is valid for a chemical reaction that proceeds
along the reaction coordinate over the transition state is extended to molecular
dynamics trajectories that in general do not cross the dividing surface between the
initial and the final local minima at the exact transition state. Our molecular dynamics
Bell-Evans-Polanyi principle
states that low energy molecular dynamics trajectories are more likely to
lead into the basin of attraction of a low energy local minimum than high energy trajectories.
In the context of global optimization schemes based on molecular dynamics
our molecular dynamics Bell-Evans-Polanyi principle implies that using low energy trajectories
one needs to visit a smaller number of distinguishable local minima before finding
the global minimum than when using high energy trajectories.
\end{abstract}

\pacs{PACS numbers: 71.15.-m, 71.15.Mb }

\maketitle
\begin{figure}[h]             
\begin{center}
\setlength{\unitlength}{1cm}
\begin{picture}( 11.2,6.4)           
\put(-1.22,-1.1){\includegraphics{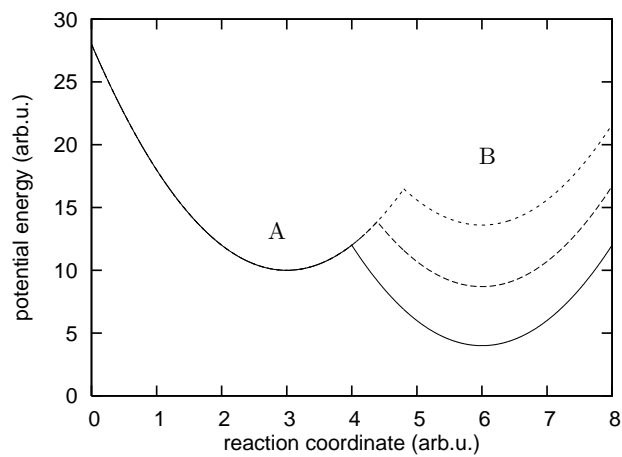}}   
\put(3.6,3.0){A}   
\put(6.4,4.0){B}   
\end{picture}
\caption{ \label{BEPcols} The potential energy surface in the region between minimum A and 
B in the Marcus approximation. The potential energy in the neighborhood of minimum A is given by a 
parabola centered at A, and in the neighborhood of minimum B by a parabola centered at B. 
The activation energy ${\epsilon}_a$ for the chemical reaction $A\rightarrow B$ is determined by the 
intersection of the two parabolas. It can be seen that the activation energies are smaller  if the 
local minimum B is lower in energy. }
\end{center}
\end{figure}

The Bell-Evans-Polanyi (BEP) principle ~\cite{textbook_1,textbook_2} is an important
fundamental principle in chemistry.
It gives a relation between the free energy $\Delta G$ released in a chemical reaction
and the activation free energy ${\epsilon}_a$ for the reaction. It was qualitatively first put forward by
Br$\o$nsted \cite{broensted} who observed that strongly exothermic reactions have a low activation energy.
A more quantitative relation was then derived by Polanyi {\mbox {\it et al}\cite{bep_pr,textbook_1}} who
approximated the potential energy surface by straight lines. This approximation leads to a linear
relation between the activation energy  ${\epsilon}_a$ and the free energy of the reaction $\Delta G$:
\begin{equation} \label{bep_lin}
 {\epsilon}_a=k_1+k_2\Delta G \quad ,
\end{equation}
where $k_1$ and $k_2$ are constants that depend on the slopes of the lines.
A more accurate approach by Marcus\cite{Marcus,textbook_2} approximates the potential energy surface
by two parabolas centered at the two local minima of the energy. The potential energy surface
in this approximation is everywhere a quadratic form with the exception of the
intersection point of the two parabolas where it has a discontinuity in its derivative.
From Fig.\ref{BEPcols} it is easily visible that the barrier for the reaction $A\rightarrow B$ is lowered if
the parabola centered in the minimum B is shifted downward. The resulting  quantitative relation\cite{textbook_2} is given by
\begin{equation} \label{bep_quad}
{\epsilon}_a=k+\frac{\Delta G}{2}+\frac{\Delta G^2}{16k}\quad ,
\end{equation}
where $k$ is proportional to the curvature of the two parabolas.

In a chemical reaction the reaction coordinate connects the educt A with the product B. 
Hence the intersection of the two parabolas is the transition state. In this article we 
will study the BEP principle not for this hypothetical path along 
the reaction coordinate but for {\mbox {molecular dynamics}} (MD) trajectories 
that cross the dividing hypersurface between the two basins of attraction~\cite{en_land_wales}
of two local minima  
on the potential energy surface. The notions of educt and product are replaced by the notions 
of initial and final local minima in this context. We will show that the BEP principle is also valid in the 
context of MD. 

\begin{figure}[h]             
\begin{center}
\setlength{\unitlength}{1cm}
\begin{picture}( 11.2,6.0)           
\put(-0.4,-0.6){\includegraphics{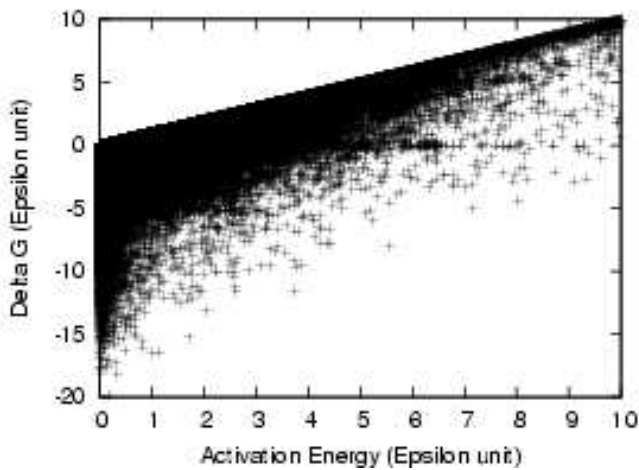}}   
\end{picture}
\caption{ \label{bepscatter} The relation between the activation energy $G^s-G^a$ and 
the reaction energy $G^b_i- G^a_i$ for more than 130000 saddle points in a Lennard Jones cluster of 55 atoms.
All the energies plotted here are free energies at T = 0, i.e. just energies }
\end{center}
\end{figure}

\begin{figure}[h]             
\begin{center}
\setlength{\unitlength}{1cm}
\begin{picture}( 11.2,6.0)           
\put(-0.4,-0.6){\includegraphics{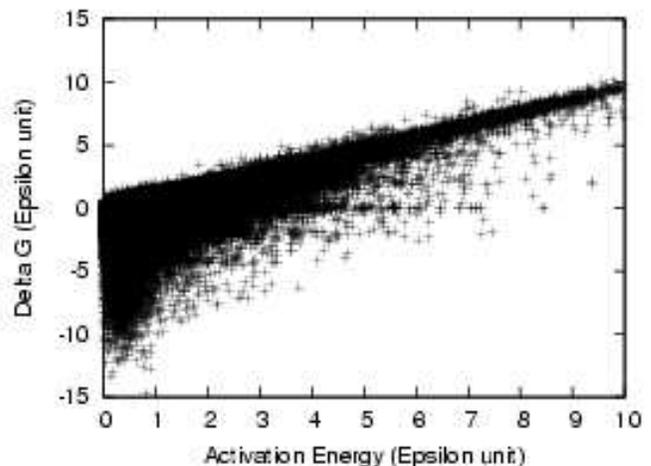}}   
\end{picture}
\caption{ \label{bepscatter_f} The relation between the activation energy $G^s-G^a$ and
the reaction energy $G^b_i- G^a_i$ for more than  130000 saddle points  at T = 30 K.}
\end{center}
\end{figure}

\begin{figure}[h]             
\begin{center}
\setlength{\unitlength}{1cm}
\begin{picture}( 11.2,4.94)           
\put(-1.24,-1.44){\includegraphics{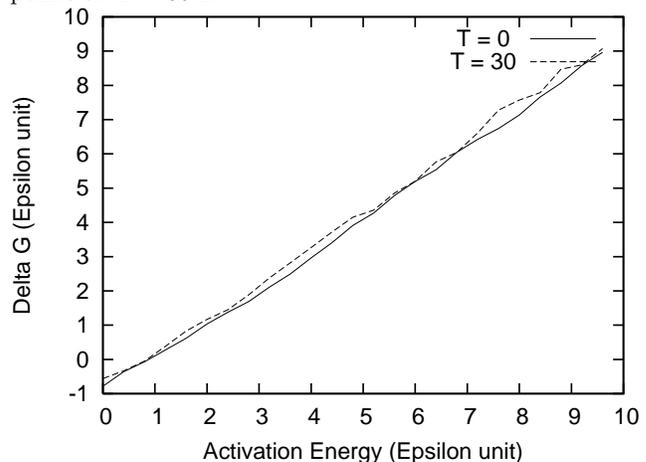}}   
\end{picture}
\caption{ \label{bepbins} The same data as in Fig.~\ref{bepscatter} and  Fig.~\ref{bepscatter_f} but averaged 
within 25 bins along the x axis.}
\end{center}
\end{figure}

Since our study requires the calculation and {\mbox statistical} evaluation of a very large number 
of local minima and saddle points, we will initially base our study on a Lennard Jones 
cluster~\cite{wales},~\cite{en_land_wales} containing 
55 atoms for which stationary points can be calculated rapidly. 
The parameters entering in the Lennard Jones potential were selected such that it models Argon clusters, namely 
$\epsilon$=0.998 kJ/Mole, $\sigma$=3.4$\AA$ and M=39.948amu~\cite{en_land_wales}.
The free energies were calculated within the harmonic approximation as the vibrational free energy. 
The temperature at which the free energies were calculated is 30 K which is below the melting point (50 K) 
of this weakly bound system~\cite{en_land_wales}(Lennard Jones cluster).
Initially we have searched for more than  130000 first order saddle points $G^s_i$ on the potential energy surface connecting energetically low local minima. Subsequently we have moved the system to the left and to the right along the direction where the curvature is negative. These two points served as the starting points for a local geometry optimization that led us in the two closest local minima. In this way we have generated 
pairs of local minima together with the saddle points that connect them.
Fig.~\ref{bepscatter_f}  and Fig.~\ref{bepscatter} show  scatter plots 
of $\Delta G=G^b_i- G^a_i$ versus the activation energy $\epsilon_a=G^s_i-G^a_i$ with and  without the entropy  
contributions respectively  and Fig.~\ref{bepbins} shows a histogram with averages of the $G^s_i-G^a_i$. 
Each pair of local minima contributed two data points to these plots since one can surmount the barrier by going 
from the minimum A to minimum B as well as by going from minimum B to minimum A.

The two scatter plots in Fig.~\ref{bepscatter} and  Fig.~\ref{bepscatter_f}
show that there is no strict linear correlation 
between the barrier height $\epsilon_a$ and the energy difference $\Delta G$ between the two minima.
For small barrier heights one can find both high energy and low energy minima behind 
the barrier. However, the BEP principle holds as a negation.
If one goes over high barriers it is extremely unlikely that one will end up in a low energy minimum. 
The better correlation for large activation energies is simply due to the fact that $\Delta G$ can not become larger than $\epsilon_a$. 
On the other hand, Fig.~\ref{bepbins} shows that there is a good linear relation if one averages over $\Delta G$.
Good linear Bell-Evans-Polanyi relations have been found in calculations of dissociative chemisorption of various molecules~\cite{bep_ex_1,bep_ex_2,bep_ex_3,cat_bep}.

\begin{figure}[h]             
\begin{center}
\setlength{\unitlength}{1cm}
\begin{picture}( 11.0,5.9)           
\put(-1.53,-1.5){\includegraphics{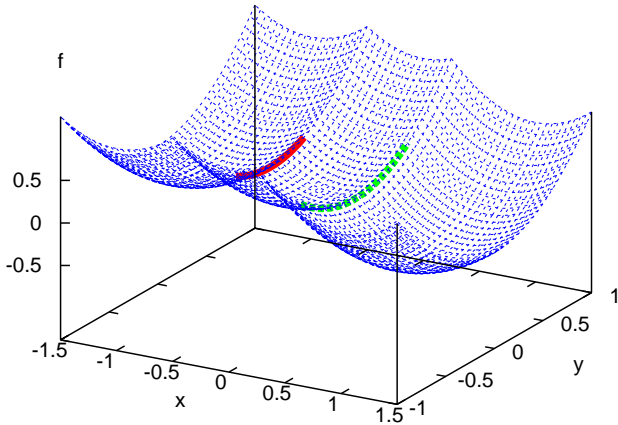}}   
\end{picture}
\caption{ \label{bepcross} This figure shows in blue the potential 
energy surface of 3 local minima together with their 
basins of attraction. The value at the minimum is 0 for the minimum in the 
middle and in the background, but -0.5 for the minimum in the foreground.
The region where a MD trajectory of energy .25 can cross from the central 
basin of attraction into the foremost basin of attraction (green) is much wider 
than the region where the trajectory can cross into the basin in the background (red). 
The value of the partition function $Q_s$ for crossings into the basin in the 
foreground is thus expected to be much larger than for crossings into the 
basin in the background.}
\end{center}
\end{figure}

Kinetic rate theory gives the rate constant for 
a reaction as
$$ k = \frac{k_B T}{h} exp(-\beta \epsilon_a)  = \frac{k_B T}{h} \frac{Q_s}{Q_a} exp(\beta (E^a-E^s))  \quad ,$$
where $E_a$ and $E_b$ are the energies of the two minima. 
$Q_s$ is the partition function for the transition state and gives in 
a certain sense the size of the barrier region. 
The importance of entropy terms can easily be seen in the classical limit. 
By making the same approximation as was 
done originally by Marcus in the derivation of the BEP principle, namely that the potential 
energy surface is the union of parabolas, but by considering {\mbox 2-dimensional} parabolas 
instead of {\mbox 1-dimensional} parabolas, one can easily 
see in Fig.~\ref{bepcross} that the size of the crossing surface that can be surmounted 
by a MD path of limited energy is increasing 
strongly when the MD path goes into an energetically low basin. We expect therefore that 
for a fixed energy crossings into low energy minima are more probable than crossings into high energy 
local minima. 

\begin{figure}[h]             
\begin{center}
\setlength{\unitlength}{1cm}
\begin{picture}( 10.0,5.8)           
\put(-1.55,-1.5){\includegraphics{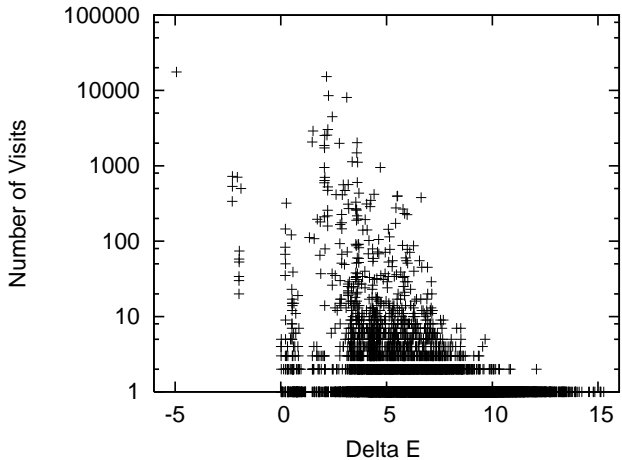}}   
\end{picture}
\caption{ \label{bepvisits} The number of visits as a function of $E_b-E_a$.}
\end{center}
\end{figure}

Fig.~\ref{bepvisits} shows the results of a numerical experiment. For MD trajectories 
that start with random directions but fixed kinetic 
energy $E_{kin}$ from a certain minimum with energy $E_a$ we have recorded how many 
times this 
trajectory reaches the basin of attraction of neighboring minima with energy $E_b$. 
To check whether the MD trajectory has crossed into another basin of attraction 
steepest descent geometry optimizations were started after every 20 MD steps. 
In Fig.~\ref{bepvisits} we then plot the number of visits as a function of 
$E_b-E_a$. We see that it is orders of magnitude more likely that the MD trajectory 
crosses into low energy basins than in high 
energy basins. We will denote this correlation as the MDBEP principle: 
low energy MD trajectories are more likely to
lead into the basin of attraction of a low energy local minimum than high energy trajectories.
The activation energy of the original BEP principle has thus been replaced by the energy of 
the trajectory. This implies that we have replaced of property of the potential energy surface 
by a property of the MD trajectory exploring this surface.

\begin{figure}[h]             
\begin{center}
\setlength{\unitlength}{1cm}
\begin{picture}( 10.0,5.8)           
\put(-1.2,-1.35){\includegraphics{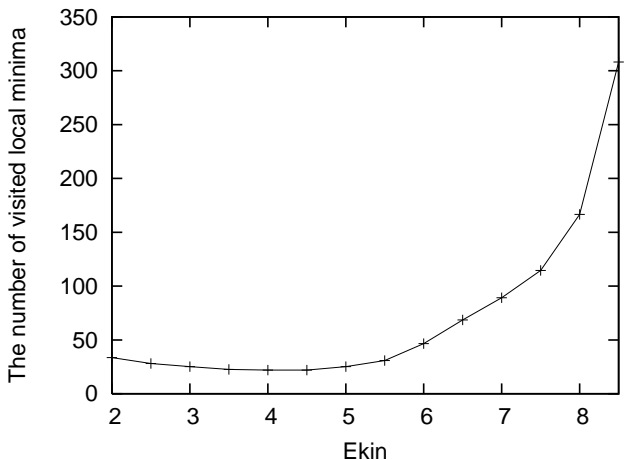}}   
\end{picture}
\caption{ \label{beplj55}The MDBEP principle for the Lennard-Jones cluster of 55 atoms.}
\end{center}
\end{figure}

As can be seen from Fig.~\ref{bepscatter}, Fig.~\ref{bepscatter_f} and Fig.\ref{bepvisits}, both the traditional BEP principle 
and our MDBEP principle are only valid in an average sense. Such a validity on the 
average is sufficient in the context of global optimization using the minima 
hopping method (MHM)~\cite{minhop,DMHM}.  In MHM the system moves from 
one local minimum to another by a combination of MD followed by a 
local geometry optimization. With the MD part one jumps from one minimum into the basin 
of attraction of another minimum. The subsequent local geometry optimization part brings us 
then into the local minimum of this basin of attraction. 
In the {\mbox original publication ~\cite{minhop}} it was already 
pointed out that the BEP principle is at least partly responsible for the success 
of the minima hopping method. If the MD trajectory has 
a small kinetic energy $E_{kin}$, it can not go over very high barriers and it is thus 
more likely to reach the basins of attraction of low lying minima. It was 
shown that the number of local minima that was visited before the global minimum was 
found decreases when the kinetic energy $E_{kin}$ of the trajectory is reduced. 
Fig.~\ref{beplj55} demonstrates the MDBEP principle for the Lennard-Jones cluster 
consisting of 55 atoms. There is a very strong correlation between the energy 
of the MD trajectory and the number of minima that are visited before the global 
minimum is found.
The data for Fig.~\ref{beplj55} and the following similar figure were obtained by 
performing MHM runs that are stopped once the global minimum is found for 
different but fixed kinetic energies $E_{kin}$ (i.e. $\beta_1=\beta_2=\beta_3=1$ 
using the notation of ref.~\cite{minhop}) in a 
reasonably chosen energy interval. Subsequently we plot the values of $E_{kin}$ versus 
the number of local minima that were visited before the global minimum was found. The potential 
energy of the local minimum from which the MD trajectory starts is set to zero. In this 
way the kinetic energy is the total energy of the MD trajectory and by energetic 
reasons it can not cross barriers higher than $E_{kin}$ relative the starting minimum. 
Only new and accepted local minima are counted. In order to achieve better {\mbox statistics} 
we perform for each fixed $E_{kin}$ 100 MHM runs (for Fig.~\ref{beplj55} the average is taken over 1000 runs), and take for the plots the 
{\mbox {averaged}} number of visited local minima. 

Since the Lennard Jones potential is a drastic simplification of the 
true inter-atomic interactions in solid state systems one might wonder whether 
the MDBEP principle also holds true for more realistic interactions. 
Using the MHM, we will therefore examine in the following the validity of the MDBEP 
principle for other systems, namely Morse clusters and 
silicon clusters described both by a force field and a tight binding scheme. 

\begin{figure}[h]             
\begin{center}
\setlength{\unitlength}{1cm}
\begin{picture}( 10.0,5.8)           
\put(-1.2,-1.35){\includegraphics{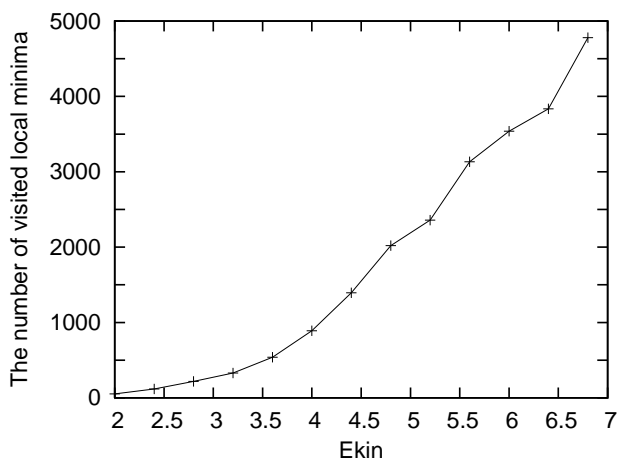}}   
\end{picture}
\caption{ \label{bepmorse6}The MDBEP principle for the Morse cluster cluster of 38 atoms with $\rho=6.0$}
\end{center}
\end{figure}

\begin{figure}[h]             
\begin{center}
\setlength{\unitlength}{1cm}
\begin{picture}( 10.0,5.8)           
\put(-1.2,-1.35){\includegraphics{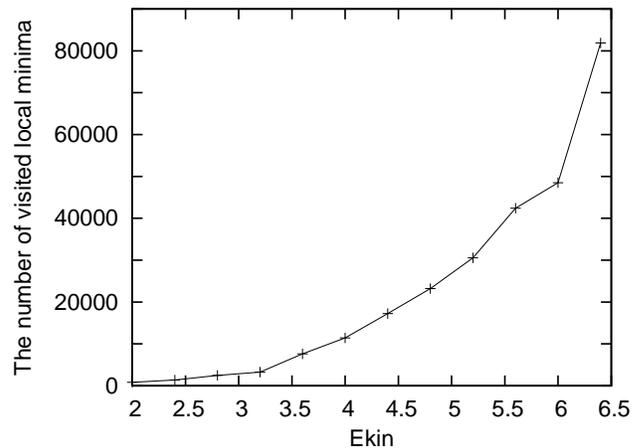}}   
\end{picture}
\caption{ \label{bepmorse10}The MDBEP principle for the Morse cluster of 38 atoms with $\rho=10.0$}
\end{center}
\end{figure}

Fig.~\ref{bepmorse6} and Fig.~\ref{bepmorse10} present our results for 
the Morse clusters of 38 atoms with $\rho=6.0$ and $\rho=10.0$. Large values of $\rho$
lead to a interaction that varies over shorter length scales. As a consequence the 
potential energy surface becomes more rugged and has significantly more local minima~\cite{{en_land_wales}}. 
As a consequence considerably more minima are visited before the global minimum is found. 
The MDBEP principle is however well observed in both cases. 

\begin{figure}[h]             
\begin{center}
\setlength{\unitlength}{1cm}
\begin{picture}( 10.0,5.9)           
\put(-1.2,-1.35){\includegraphics{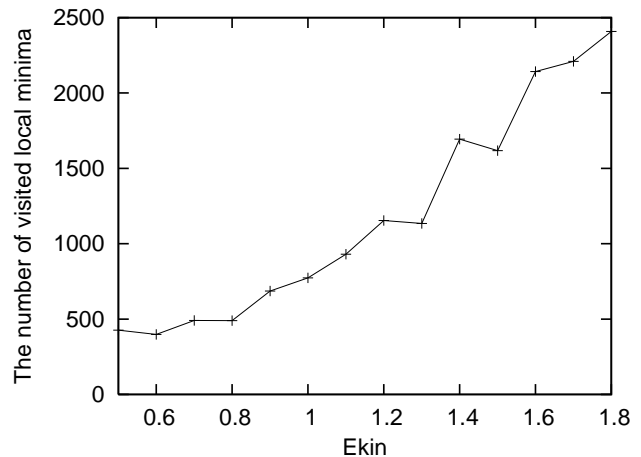}}   
\end{picture}
\caption{ \label{beptb20}The MDBEP principle for the Lenosky tight binding cluster of 20 atoms.}
\end{center}
\end{figure}

\begin{figure}[h]             
\begin{center}
\setlength{\unitlength}{1cm}
\begin{picture}( 10.0,5.8)           
\put(-1.2,-1.35){\includegraphics{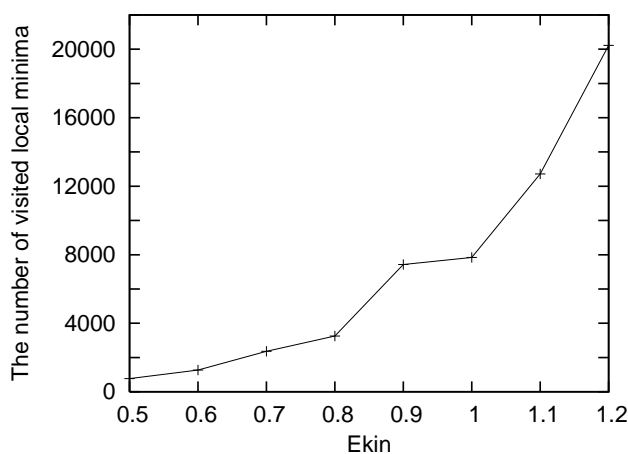}}   
\end{picture}
\caption{ \label{bepff33}  The MDBEP principle for the Lenosky force field cluster of 33 atoms.}
\end{center}
\end{figure}

Fig.~\ref{beptb20} and Fig.~\ref{bepff33} present our results for the Si$_{20}$ cluster
within the Lenosky tight binding scheme ~\cite{Len_TB} and for the Si$_{33}$ cluster
within the Lenosky force field~\cite{Len_FF}. In contrast to the Lennard Jones and Morse 
potentials the silicon force field has much more complicated interactions that 
depend not only on the distance between atoms but also on things like the bond angles. 
Tight binding schemes are the simplest way to treat solid state systems at a quantum mechanical 
level.  The Lenosky tight binding scheme gave a very good 
agreement with the DFT energies~\cite{DMHM} and can be considered as a reliable approximation to 
a precise density functional treatment of silicon clusters. 
The MDBEP principle is valid in both cases which demonstrates that the MDBEP principle 
is also valid for realistic interactions and in particular for quantum mechanical interactions. 

The fact that for small values of $E_{kin}$ the global minimum is found after having 
visited only a small number of local minima does not imply that the computational 
time in the MHM is continuously decreasing with smaller values of $E_{kin}$.
If $E_{kin}$ is getting too small the system has to make many attempts before 
succeeding to escape from the basin of attraction of the current minimum and this 
will actually lead to an increase in the computer time. 

In practice, the shortest computation time can be obtained by giving the MD trajectories initial 
velocities that have large components in the subspace of low curvature of the Hessian matrix.
Due to the fact that low energy saddle points often lie at the end of low-curvature 
modes~\cite{dimer_method,act_relax,eigv_follow} one can in this way even with low energy 
trajectories rapidly escape from the present minimum. 

In summary, we have shown that the BEP principle can be extended to MD 
trajectories. We call this extended principle MDBEP principle and it says that MD 
trajectories with low energy are more likely to lead into basins of attraction 
of low energy configurations. Having verified this principle numerically for 
several systems we believe that it is valid for any solid state system.

We thank the Swiss National Science Foundation for the financial support of our research.

\end{document}